\documentclass[11pt]{article}
\usepackage[margin=1in]{geometry}
\usepackage{booktabs}
\usepackage{graphicx}
\usepackage{amsmath,amssymb}
\usepackage[numbers,sort&compress]{natbib}
\usepackage{float}
\usepackage{hyperref}
\hypersetup{colorlinks=true, linkcolor=blue, citecolor=blue, urlcolor=blue}

\title{Beyond SDR: How Music Source Separation Reshapes Rhythm-Relevant Signal Properties}
\author{Chuxin Ding\\ {Universitat Autònoma de Barcelona} \\ \texttt{{chuxinding@outlook.com}}}
\date{}

\begin{document}
\maketitle

\begin{abstract}
Music source separation (MSS) is increasingly used not to remix music but to \emph{measure} it: separated drum stems feed studies of microtiming, dynamics, and groove. The field evaluates separators almost exclusively by signal-to-distortion ratio (SDR), yet microrhythm research shows that a sound's perceived temporal location (its p-centre) is co-determined by its attack and envelope, precisely the properties SDR was not designed to protect. We quantify what four open separators spanning four architecture generations (Spleeter, HT-Demucs, BS-Roformer, SCNet-XL) do to rhythm-critical signal properties, using the 50-track MUSDB18-HQ test set, where true stems make every claim falsifiable. Three findings emerge. (1)~Onset \emph{timing} is safe: onset F-measure tracks SI-SDR (Spearman $\rho = 0.62$) and is invariant to input length. (2)~Transient and dynamic shape are not: their distortion correlates only weakly with SI-SDR ($|\rho| \leq 0.29$), and the model ranking inverts---the SDR leader distorts drum attacks
twice as much as its capability-matched CNN counterpart, while the SDR-worst model preserves dynamics better than a mid-pack one. Each model imposes a systematic, model-specific bias on the dynamic profile. (3)~Input length reshapes the rendered attack of a fixed passage (marginally more for the transformer; paired $p = 0.044$) while leaving onset locations untouched. For rhythmic studies, separator choice and input conditions are methodological variables to be reported, and SDR alone cannot stand in for them.
\end{abstract}

\section{Introduction}

Musicological studies increasingly rely on music source separation as an instrument. To measure a drummer's microtiming, dynamics, and timbre inside a dense mix, the analyst first isolates the drum stem with a neural source-separation model, then measures the stem as if it were the performance. The validity of every downstream number therefore rests on an assumption that is rarely tested: \emph{that separation largely preserves the signal properties the measurement reads.}

The separation field evaluates progress with the signal-to-distortion ratio (SDR) family \citep{vincent2006,leroux2019}, the figure of merit of the MDX/SDX challenge series \citep{mitsufuji2022,fabbro2024}. SDR measures global reconstruction fidelity. But it aligns imperfectly with perceptual quality \citep{torcoli2021}, and---more specifically for rhythm research---it says nothing about \emph{which} properties of the signal absorb the distortion. This matters because of a robust finding in microrhythm research: the perceived temporal location of a sound, its perceptual centre or \emph{p-centre}, is co-determined by the sound's attack, duration, and envelope, not by its physical onset alone \citep{morton1976,danielsen2019,danielsen2024}. A separator that reshapes a drum hit's attack does not merely add measurement noise; it can shift the microrhythmic relationship itself while leaving the physical onset, and the SDR, essentially intact.

Qualitative groundwork for this concern was laid in the companion study \citep[\S4.7]{ding2026}, which compared separators by close listening and spectrographic inspection and observed model-specific signatures: systematic reshaping of the drum stem's dynamic profile, transient smearing and premature apparent onsets in frequency-domain models, and a dependence of output character on the \emph{length} of the input audio. The present paper turns those observations into falsifiable claims and tests them.

\paragraph{Contributions.}
\begin{enumerate}
\item A two-family evaluation protocol that scores separators simultaneously with the field's \emph{developer} metrics (BSS-Eval SDR/SIR/SAR, SI-SDR) and with \emph{analyst} metrics that read part of what a musicologist tend to read: onset timing fidelity, an eight-descriptor
dynamic profile measured against the true stem, and ground-truth energy-routing (labeling) precision.
\item A benchmark of four open separators spanning 2019--2024 architecture generations on MUSDB18-HQ, showing that SDR predicts timing fidelity but not transient or dynamic fidelity, with concrete ranking inversions at the state-of-the-art frontier (\textbf{B3}), and that each model leaves a systematic, model-specific bias on the dynamic profile (\textbf{A1}).
\item A controlled input-length experiment showing that the amount of audio surrounding a fixed passage changes how its transients are rendered---most for the attention-based model---while its onset locations remain stable (\textbf{A2}).
\item An open, cached, resumable pipeline (configs, runners, metrics, synthesis, and all result tables) for reproducing and extending every number in this paper.
\end{enumerate}

\section{Background and hypotheses}

\subsection{Separation models and their evaluation}

Modern MSS models fall into frequency-domain, time-domain, and hybrid families. Spleeter \citep{hennequin2020}, a 2019 spectrogram U-Net, became the first widely deployed open baseline. Demucs evolved from a waveform U-Net into the hybrid HT-Demucs, which fuses temporal and spectral branches through a cross-domain transformer \citep{defossez2021,rouard2023}. Band-split architectures then pushed the state of the art: BS-Roformer applies rotary-position transformers over learned sub-bands \citep{lu2024}, while SCNet reaches comparable quality with a sparse-compression CNN and no attention \citep{tong2024}. Throughout, the reported quantity is SDR --- its framewise BSS-Eval form and the challenges' utterance-level variant (uSDR: one global SDR per track, averaged over tracks) --- on MUSDB18 \citep{rafii2017,stoter2019} or its challenge variants; perceptual validation is sparse, and the top of recent leaderboards is compressed, with the three best SDX'23 systems sitting within $0.79$~dB average SDR \citep{fabbro2024} with no clear perceptual ordering \citep[cf.][]{torcoli2021}.

\subsection{Why rhythm research needs more than SDR}

Rhythm and groove studies read three things off a drum stem: \emph{when} events occur (onset/microtiming), \emph{how strongly} (dynamics), and \emph{with what envelope and timbre} (which, via the p-centre, feeds back into perceived timing). The p-centre literature shows that attack sharpness and envelope shape displace a sound's perceived temporal location by tens of milliseconds \citep{danielsen2019}, the same order of magnitude as the microtiming deviations such studies quantify. Distortion concentrated in the attack is therefore \emph{qualitatively} different from distortion spread across the spectrum, and a scalar reconstruction metric cannot distinguish the two.

\subsection{From qualitative observation to falsifiable claims}

From the companion study's observations we derive three claims, each testable on ground truth:
\begin{itemize}
\item \textbf{A1 (model-specific dynamics bias).} The dynamic profile of a separated stem deviates from the true stem in a way that depends systematically on the separator. If true, a dynamics statistic computed through separation is confounded by model choice.
\item \textbf{A2 (input-length sensitivity).} Separating the \emph{same} passage inside different amounts of surrounding audio changes its rendered transients, dynamics and labeling. A mechanistic corollary: an attention model, able to exploit long-range structure inside its input window, should be more context-sensitive than a frame-local CNN.
\item \textbf{B3 (SDR dissociation).} Ranking models by SDR and by analyst-relevant fidelity are different rankings; pooled correlation between SDR and analyst distortion is weak.
\end{itemize}

\section{Materials and method}

\subsection{Models}

Four open separators spanning architecture generations, all emitting the standard four stems (drums/bass/vocals/other), all run deterministically (fixed parameters, no re-initialisation), summarized in Table~\ref{tab:models}. BS-Roformer and SCNet-XL were chosen as a \emph{capability-matched} attention/no-attention pair (their published MUSDB drums SDRs differ by $0.1$~dB), so that architecture effects are not confounded with overall quality. Proprietary tools examined qualitatively in the companion study have no scriptable open implementation and are excluded here.

\begin{table}[h]\centering
\caption{The four separators. All are 4-stem; all run deterministically.}
\label{tab:models}
\begin{tabular}{llll}
\toprule
Model & Year & Family & Runner \\
\midrule
Spleeter & 2019 & frequency-domain CNN (U-Net) & \texttt{spleeter:4stems} v2.4.2 (TF, CPU) \\
HT-Demucs & 2022 & hybrid waveform+spectrogram+transformer & \texttt{htdemucs} v4 (PyTorch, GPU) \\
BS-Roformer & 2023 & band-split rotary transformer & MSST ckpt, drums 11.6 dB (GPU) \\
SCNet-XL & 2024 & sparse-compression CNN & MSST ckpt, drums 11.5 dB (GPU) \\
\bottomrule
\end{tabular}
\end{table}

\subsection{Data}

The MUSDB18-HQ \emph{test} subset: 50 full tracks, 44.1~kHz stereo WAV, each with true isolated stems. Every (model, track) separation is cached to disk by default, making all experiments resumable and re-scorable without re-separation.

\subsection{Metrics}

\paragraph{Developer family.}
BSS-Eval v4 SDR/SIR/SAR/ISR (framewise, 1~s windows, median over frames; via \texttt{museval}) and scale-invariant SDR \citep{leroux2019}. These are the numbers vendors and challenges report, recomputed on our tracks as the comparison baseline.

\paragraph{Analyst family.} Computed per stem against the true stem:
\begin{itemize}
\item \emph{Timing} --- onset F-measure: onsets detected identically on estimate and reference (spectral flux with backtracking; \texttt{librosa} \citep{mcfee2015}), matched within $\pm 50$~ms (\texttt{mir\_eval} \citep{raffel2014}). Precision, recall, and F-measure (F, the harmonic mean of precision and recall).
\item \emph{Dynamics} --- an eight-descriptor \textbf{dynamic profile} from a root-mean-square (RMS) amplitude envelope (20~ms analysis frames, 10~ms hop; 50\% overlap): level (dB), variation (SD of the dB envelope), crest, impulsiveness, envelope skewness, envelope flux, attack slope (mean rising-edge steepness), and burst rate. Each descriptor is reported raw on the estimate (\texttt{dyn\_*}) and as a signed deviation from the true stem (\texttt{d\_* = estimate $-$ truth}); $|\texttt{d\_*}|$ is a distortion magnitude.
\item \emph{Labeling} --- each estimated stem is least-squares-decomposed onto the set of true sources (MUSDB stems sum to the mixture, so the decomposition is principled); \textbf{labeling precision} is the correctly-labelled source's share of explained energy (gain-invariant; 1.0 = no cross-routing).
\end{itemize}

\subsection{Experimental designs and statistics}

\paragraph{Benchmark (A1, B3, B1).} All 4 models $\times$ 50 tracks $\times$ 4 stems; analysis focuses on drums (the rhythm-critical stem; $n = 200$ model-track observations), with the other stems as controls. A1a compares the cross-model spread of each dynamic descriptor with the between-track spread within models; A1c asks whether models even agree on \emph{which} tracks are most distorted (between-model Spearman of per-track \texttt{d\_variation}). B3 compares model rankings under SDR versus analyst metrics and pools per-track observations for Spearman correlations of SI-SDR against each distortion magnitude.

\paragraph{Input-length sweep (A2).} For each track a fixed 5~s region \textbf{R} (default 30--35~s) is separated inside four input windows that all contain R: R alone, 15~s, 30~s, and the full track. The output aligned to R is extracted and measured on R only. \textbf{Drift} is the standard deviation of a descriptor across the four windows, per (model, track). Run on 20 tracks with the three GPU models (Spleeter is architecturally the oldest and CPU-bound; it is not required for the attention contrast). The attention hypothesis is tested with paired Wilcoxon signed-rank on per-track drift, BS-Roformer vs.\ SCNet-XL. Because ground truth exists, we also ask whether growing context moves R \emph{toward} the true stem ($|\texttt{d\_*}|$ shrinking) or merely changes it.

\subsection{Reproducibility}

All code, configuration, result tables, and figures are in the companion repository (\url{https://github.com/dingchuxin/MSS-eval}).
Separation environments are isolated per dependency stack (PyTorch 2.11/CUDA 12.8 for the GPU models; TensorFlow 2.12/CPU for Spleeter); every separation carries provenance metadata \{model, version, parameters, input range, sample rate\}. GPU: one NVIDIA RTX 4070 SUPER (12~GB).

\section{Results}

\subsection{The developer view: a clean generational story}

Median BSS-Eval SDR per stem (Table~\ref{tab:sdr}) reproduces the published picture---a monotonic climb across generations and a compressed frontier. On drums the two 2023/24 models are separated by $0.05$~dB; by the field's own metric they are interchangeable. Labeling precision is near-ceiling for all models and stems (medians 0.94--1.00): modern separators route energy to the right stem label; energy is the open question.

\begin{table}[h]\centering
\caption{Median BSS-Eval SDR (dB) per stem, 50 tracks.}
\label{tab:sdr}
\begin{tabular}{lrrrr}
\toprule
Model & bass & drums & other & vocals \\
\midrule
Spleeter & 4.96 & 5.69 & 4.22 & 6.35 \\
HT-Demucs & 9.78 & 10.06 & 6.51 & 8.63 \\
SCNet-XL & 10.87 & 11.63 & 8.00 & 10.87 \\
BS-Roformer & 9.68 & 11.68 & 8.12 & 10.78 \\
\bottomrule
\end{tabular}
\end{table}

\subsection{A1: the dynamic profile carries a systematic, model-specific bias}

Every dynamic descriptor deviates from truth in a way that differs \emph{systematically by model}. The clearest case is dynamic variation: median $|\texttt{d\_variation}|$ on drums is \textbf{1.78~dB for BS-Roformer but 4.09~dB for HT-Demucs}. HT-Demucs reshapes the drum dynamic envelope $\sim$2.3$\times$ more, a direct quantitative confirmation of the companion study's observation that HT-Demucs compresses dynamics. The bias is not a proxy for overall quality: HT-Demucs (10.1~dB SDR) distorts dynamics more than Spleeter (5.0~dB SDR; $|\texttt{d\_variation}|$ 2.41~dB).

Two qualifications worth noting. First, the between-model spread is \emph{smaller} than the between-track spread: across the eight deviation descriptors, the ratio of cross-model SD to mean within-model (between-track) SD ranges 0.05--0.56 (0.36 for \texttt{d\_variation}). Model choice is a real but secondary variance source, a systematic per-model \emph{offset}, not the dominant source of scatter. Second, models only moderately agree on which tracks are hardest: the between-model Spearman correlation of per-track \texttt{d\_variation} averages \textbf{0.49} (range 0.23--0.60), so even the track-level ranking of distortion is partly model-dependent.

The consequence for practice stands regardless: because the bias is systematic and model-specific, dynamics measured through different separators are not comparable across studies, and no global correction removes the offset.

\subsection{B3: SDR predicts timing, but not transient or dynamic shape}

\begin{figure}[H]\centering
\includegraphics[width=\linewidth]{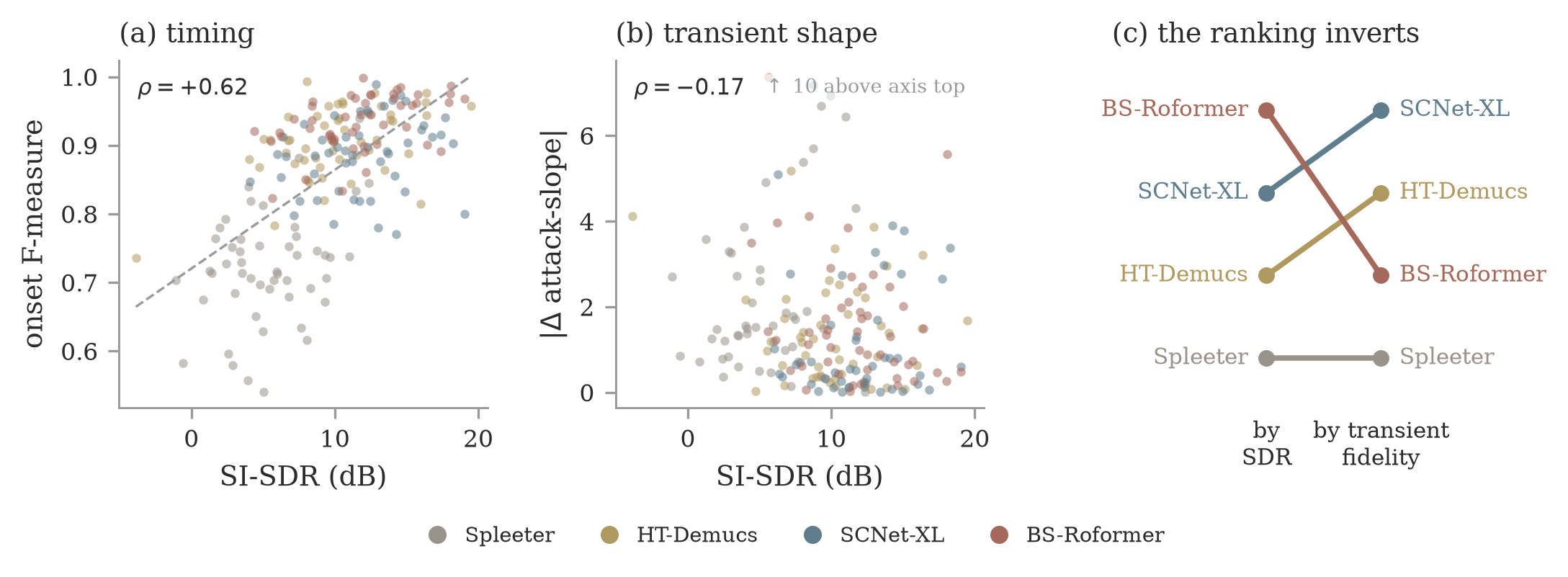}
\caption{SI-SDR vs.\ analyst fidelity on the drums stem (4 models $\times$ 50 tracks). (a): onset F-measure rises with SI-SDR. (b): attack-slope distortion shows no trend against SI-SDR.}
\label{fig:sdr}
\end{figure}

\textbf{Timing tracks SDR.} Pooling all 200 drums observations, onset F-measure correlates with SI-SDR at Spearman $\rho = \mathbf{+0.62}$ ($p < 10^{-4}$). For \emph{where} onsets land, SDR is a reasonable guide (Fig.~\ref{fig:sdr} (a)).

\textbf{Shape does not.} The same pooled correlations of SI-SDR against distortion magnitudes are weak: $|\texttt{d\_attack\_slope}|$ $\rho = \mathbf{-0.17}$ ($p = 0.017$), $|\texttt{d\_variation}|$ $\rho = \mathbf{-0.16}$ ($p = 0.028$), $|\texttt{d\_crest}|$ $\rho = \mathbf{-0.29}$ ($p < 10^{-4}$). All carry the expected sign---higher SDR buys marginally less shape distortion---but the effect is far too weak for SDR to stand in for these properties (Fig.~\ref{fig:sdr} (b)).

The model ranking (Table~\ref{tab:b3}; parenthetical numbers are ranks) makes the dissociation concrete. Two inversions are diagnostic. The SDR leader \textbf{BS-Roformer distorts drum attacks more than twice as much as SCNet-XL} (1.33 vs.\ 0.64) although SDR calls the pair equal---the transient-fidelity ranking flips at the state-of-the-art frontier, exactly where leaderboard differences are smallest. And \textbf{Spleeter, last by 5--7~dB of SDR, preserves the dynamic profile better than HT-Demucs.} A researcher choosing a separator ``by the leaderboard'' would pick the model with the \emph{worst} dynamics fidelity of the modern three for a dynamics study.

\begin{table}[t]\centering
\caption{Drums, medians over 50 tracks. Parentheses give the rank on that column (1 = best, i.e.\ highest SI-SDR/onset-F or lowest $|\Delta|$).}
\label{tab:b3}
\begin{tabular}{lrrrr}
\toprule
Model & SI-SDR (dB) & onset-F & $|\Delta$ attack-slope$|$ & $|\Delta$ variation$|$ \\
\midrule
BS-Roformer & 11.88 (1) & 0.93 & 1.33 (3) & 1.78 (1) \\
SCNet-XL & 11.74 (2) & 0.89 & 0.64 (1) & 2.45 (3) \\
HT-Demucs & 10.07 (3) & 0.91 & 1.27 (2) & 4.09 (4) \\
Spleeter & 5.00 (4) & 0.72 & 1.54 (4) & 2.41 (2) \\
\bottomrule
\end{tabular}
\end{table}

\paragraph{Interpretation.} SDR is a faithful proxy for onset placement and a poor one for the shape of the attack and envelope. Via the p-centre, that is precisely the split that matters: a separator can post state-of-the-art SDR and healthy onset detection while reshaping a drum's attack enough to shift its \emph{perceived} timing, a displacement invisible to both SDR and onset F-measure. \citet{torcoli2021} reached the analogous conclusion for overall perceptual quality; we localize it to the specific signal properties rhythm research depends on.

\subsection{A2: input length reshapes transients, not timing}

\begin{figure}[H]\centering
\includegraphics[width=\linewidth]{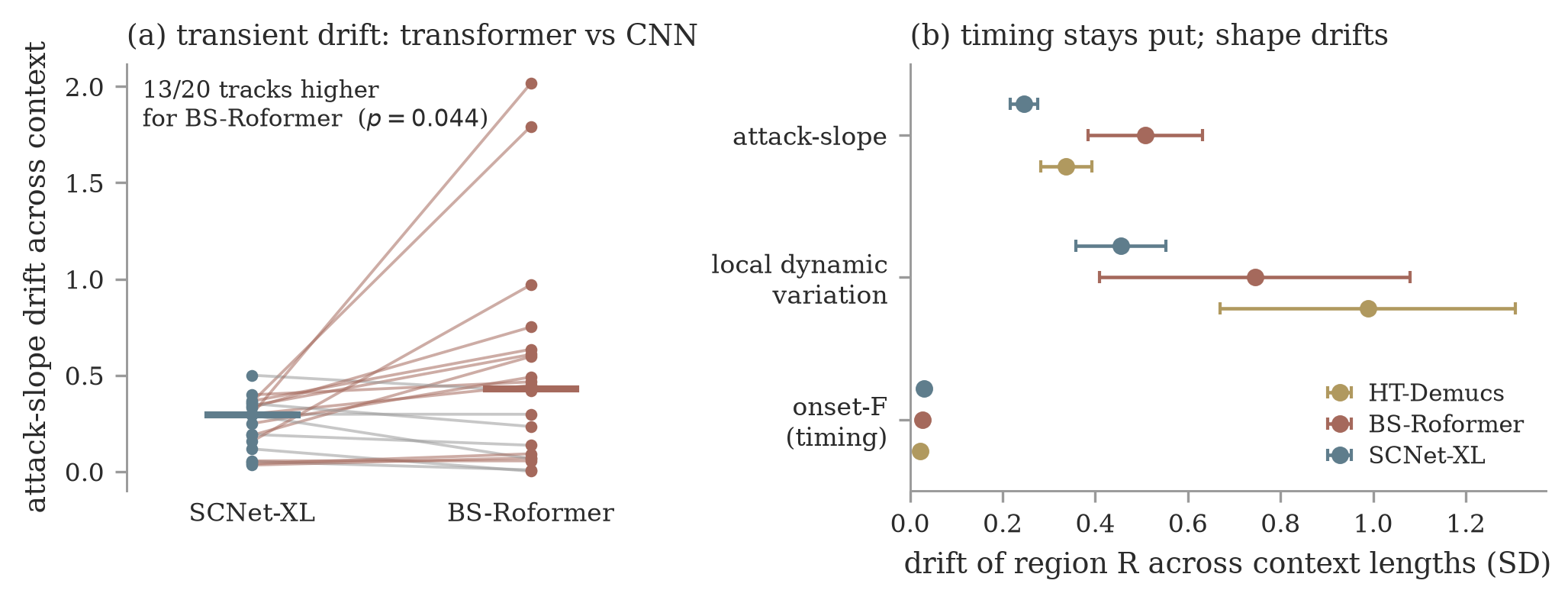}
\caption{A2 drift of the fixed drums region with input length (mean $\pm$ SE, 20 tracks). Attack-slope: transformer $>$ CNN (paired $p = 0.044$); impulsiveness: outlier-driven, n.s.; timing (onset-F): flat.}
\label{fig:a2}
\end{figure}

Drift of the fixed drums region R across the four context windows (Table~\ref{tab:a2}, Fig.~\ref{fig:a2}). \textbf{Timing is context-invariant} (onset-F drift 0.02--0.03 for every model): how much audio surrounds a passage does not move \emph{where} its onsets are detected. \textbf{Transient and dynamic shape drift} by an order of magnitude more, so input length is a real methodological variable for exactly the properties SDR already fails to protect.

\begin{table}[h]\centering
\caption{A2 drift (SD across the four context windows) of the fixed drums region, mean over 20 tracks.}
\label{tab:a2}
\begin{tabular}{lrrrr}
\toprule
Model & attack-slope & impulsiveness & variation & onset-F \\
\midrule
HT-Demucs & 0.34 & 0.36 & 0.99 & 0.023 \\
BS-Roformer & 0.51 & 2.64 & 0.75 & 0.027 \\
SCNet-XL & 0.25 & 1.43 & 0.46 & 0.030 \\
\bottomrule
\end{tabular}
\end{table}

\textbf{The attention hypothesis receives narrow, single-descriptor support from the data.} In the paired test, only attack-slope reaches significance: BS-Roformer's attack rendering drifts more than SCNet-XL's on 13/20 tracks, median ratio $1.62\times$ (Wilcoxon $p = 0.044$). The apparently large impulsiveness gap is not robust ($p = 0.67$): auditing the per-track values traced it almost entirely to a single track whose fixed measurement region contains only one drum onset---a content-blind region choice, not a model effect. BS-Roformer is also indistinguishable from HT-Demucs on attack drift ($p = 0.55$). The strong form of the hypothesis---attention makes separation broadly context-dependent---is not supported; its weak form---the transformer's \emph{attack} rendering is the most context-sensitive---is, marginally.

\textbf{Direction.} With ground truth we can ask whether more context helps. From R-alone to the full track, the CNN and hybrid move \emph{toward} truth (mean $|\texttt{d\_variation}|$ change $-0.38$ and $-0.35$), while the transformer moves slightly away ($+0.09$) even as its onset-F improves the most ($0.83 \to 0.88$). These deltas are small and offered as a suggestive asymmetry, not a significant effect: additional context appears to \emph{re-weight} rather than stabilize the transformer's transient rendering.

\subsection{B1: how the models differ: interference versus artifact}

Beyond \emph{what} the models distort, the developer metrics
reveal a systematic difference in \emph{how} they fail, and it maps onto architecture. Decomposing reconstruction into interference suppression (SIR) and artifact-freeness (SAR) on the drum stem (Table~\ref{tab:b1}), two patterns stand out. Spleeter is worst on both axes---the most residual bleed (lowest SIR) and by far the most processing artifacts (SAR 4.9, $\sim$5~dB below the modern models), quantitatively confirming the companion study's observations of leakage and envelope degradation in the 2019 frequency-domain baseline. Among the state of the art, the transformer and the CNN sit at opposite ends of an interference/artifact trade-off: \textbf{BS-Roformer suppresses interference most aggressively (highest SIR) but introduces more artifacts than SCNet-XL, which is the most artifact-free model (highest SAR).} Because the two models' SDRs are within $0.1$~dB, this is an \emph{architecture}-linked difference, not a quality difference---and it coheres with A2, in that the model leaning toward artifact-generating interference suppression (BS-Roformer) is also the one whose transient rendering is most perturbed by input context. We report this as a descriptive pattern; a per-track significance test is left to future work.

\begin{table}[h]\centering
\caption{Interference (SIR) vs.\ artifact (SAR), drum stem, median over 50 tracks (dB; higher is better on both).}
\label{tab:b1}
\begin{tabular}{lrr}
\toprule
Model & SIR (less residual bleed) & SAR (fewer artifacts) \\
\midrule
Spleeter & 12.6 & 4.9 \\
HT-Demucs & 16.8 & 8.9 \\
SCNet-XL & 17.8 & 11.0 \\
BS-Roformer & 18.9 & 9.8 \\
\bottomrule
\end{tabular}
\end{table}

\section{Discussion}

For rhythm and groove research integrating MSS, three practical guidelines follow.
\begin{enumerate}
\item \textbf{Onset-location studies are comparatively safe.} Onset F tracks SDR, is near-ceiling for modern models on drums (0.89--0.93), and is invariant to input length. Any strong modern separator will do, and results should transfer across them.
\item \textbf{Dynamics studies are not.} The dynamic profile carries a systematic, model-specific bias (A1) that SDR does not predict (B3) and input length partly perturbs (A2). Separated-stem dynamics should be reported \emph{with} the separator name, version, and input conditions, and, where possible, with the model's measured \texttt{d\_*} bias on ground-truth material. Cross-study comparison of dynamics obtained through different separators is not meaningful without it.
\item \textbf{Avoid choosing a separator on leaderboard SDR alone.} At the current frontier the choice is effectively free in SDR terms (0.05~dB) while differing by $2\times$ in attack fidelity. In this evaluation the CNN (SCNet-XL), not the SDR leader, is the transient-faithful choice for percussion analysis.
\end{enumerate}

\paragraph{The consistent fault line.} Across all experiments the same split recurs: the \emph{location} of rhythmic events is robust to model choice, to SDR level, to input length, while the \emph{shape} of those events (attack, envelope, dynamics) is fragile in every direction we probed. Given the p-centre evidence that shape co-determines perceived timing, the fragile quantity is part of the timing itself. This reframes separator evaluation for analytical use---the question is not ``how clean is the stem?'' but ``which measurements does this stem still support?''

\paragraph{Limitations.} The study reads signals, not listeners: we quantify distortion of perceptually-relevant properties but do not measure perception itself; a listening experiment linking measured \texttt{d\_attack} to perceived p-centre shift in separated stems is the natural next step. The corpus is MUSDB18-HQ---Western pop/rock-centric, and the same 100 training songs underlie most models' fine-tuning lineage. One checkpoint represents each architecture. Analysis centres on drums; other stems served as controls only. The A2 measurement region was fixed at 30--35~s regardless of musical content, which produced one near-silent-region outlier (caught in audit and excluded from inference); a content-aware region choice based on bars and phrases is preferable. And A2 varied \emph{neutral} context length only---the companion thesis's phrase-unit design (half/one/four musical phrases), which would separate musical coherence from raw duration, remains future work, now motivated by the marginal attack-slope effect found here.

\section{Conclusion}

Evaluated the way its developers evaluate it, music source separation looks solved to within a tenth of a decibel. Evaluated the way a rhythm analyst must use it, it is not solved but \emph{shaped}: every model leaves a systematic signature on the drum stem's dynamics and attack, properties that carry perceived timing, and neither SDR, nor onset detection, nor more input context reveals or repairs that signature. Source separation can serve as an instrument for rhythm research, but like any instrument it has a transfer function. This paper measured it for four models; the accompanying pipeline makes measuring it for the next model a one-command exercise.
\section*{Data and code availability}

All code, configurations, per-track tables, and figures: \url{https://github.com/dingchuxin/MSS_drum_fidelity_eval}. MUSDB18-HQ is available from Zenodo \citep{stoter2019}. Cached separations are regenerated deterministically by \texttt{scripts/run\_benchmark.py} and \texttt{scripts/run\_input\_length.py}; statistics and figures by \texttt{scripts/synthesize.py} and \texttt{scripts/synthesize\_a2.py}.

\bibliographystyle{plainnat}
\bibliography{references}

\end{document}